\documentclass[aps,showpacs,showkeys]{revtex4}
\usepackage{amssymb}
\usepackage{amsmath}
\usepackage{graphicx}

\def \be {{\begin{equation}}}
\def \ee {{\end{equation}}}
\def \bea {\begin{eqnarray}}
\def \eea {\end{eqnarray}}

\begin{document}

\title{Towards laser based improved experimental schemes for multiphoton e$%
^{+}$ e$^{-}$ pair production from vacuum.}
\author{ I. Ploumistakis$^{\dagger }$\renewcommand{\thefootnote}{***}\thanks{
Email: \texttt{iploumistakis@isc.tuc.gr}}, S. D. Moustaizis$^{\dagger }$%
\renewcommand{\thefootnote}{**}\thanks{
Email: \texttt{moustaiz@science.tuc.gr}}, and I. Tsohantjis$^{\ddagger }$%
\renewcommand{\thefootnote}{*}\thanks{
Email: \texttt{ioannis2@otenet.gr}}, }
\affiliation{Technical University of Crete, Department of Sciences, $^{\dagger }$%
Institute of Matter Structure and Laser Physics, $^{\ddagger }$Division of
Mathematics\\
Chania GR-73100, Crete, Greece}
\date{\today}

\begin{abstract}
Numerical estimates for pair production from vacuum in the presence of
strong electromagnetic fields are derived, for two experimental schemes :
the first concerns a laser based X-FEL and the other imitates the E144
experiment. The approximation adopted in this work is that of two level
multiphoton on resonance. Utilizing achievable values of laser beam
parameters, an enhanced production efficiency of up to 10$^{11}$ and 10$%
^{15} $ pairs can be obtained, for the two schemes respectively.
\end{abstract}

\maketitle



\section{Introduction}

Electron-positron pair production from vacuum breakdown in the presence of
strong electromagnetic fields is one of the most interesting non-linear QED
phenomena and at the same time a rich area of theoretical and experimental
investigations\cite{Klein, Schwinger, Brezin, Troup, greiner, Popov1,
Nikishov2, niki}. Recently the possibility for an experimental verification
of the phenomenon is plausible due to the rapid development of ultra-intense
laser facilities \cite{mourou, chen, tajima}. The first steps on the
theoretical treatment of the phenomenon can be found \cite{Klein}, but
Schwinger \cite{Schwinger} was the first that examined it thoroughly. By the
implementation of the proper time method Schwinger obtained the following
conditions for pair creation to occur : The invariant quantities $\mathcal{F=%
}\frac{1}{4}F_{\mu \nu }F^{\mu \nu }=-\frac{1}{2}\left( \mathcal{\vec{E}}%
^{2}-c^{2}\mathcal{\vec{B}}^{2}\right) $, $\mathcal{G=}\frac{1}{4}F_{\mu \nu
}\tilde{F}^{\mu \nu }=c\mathcal{\vec{E}\cdot \vec{B}}$, where $F_{\mu \nu }$
and $\tilde{F}_{\mu \nu }=\frac{1}{2}\epsilon _{\mu \nu \alpha \beta }$ $%
F^{\alpha \beta }$ are the electromagnetic field tensor and its dual
respectively, must be such that neither $\mathcal{F=}0$ , $\mathcal{G=}0$
nor $\mathcal{F>}0$ , $\mathcal{G=}0$. These restrictions are for example
satisfied close to the antinodes of a standing wave or at the region of a
focused laser beam. In order to have sizable effects the electric field
strength must exceed the critical value $\mathcal{E}_{c}$ $=\frac{mc^{2}}{%
e\lambda _{c}}\simeq 1.3\times 10^{18}V/m$ which corresponds to focal laser
intensities of the order of $10^{29}W/cm^{2}$ and above. Such laser
intensities can be reach in the very near future from facilities such as
[14], \cite{Hiper}, \cite{ELI} and XFEL [18]. Brezin and Itzykson \cite%
{Brezin} and soon after Popov \cite{Popov1} used multiphoton atom ionization
technics such as the imaginary time method \cite{Popov1,Ringwald,popov2}
showed that the probability of pair creation over a 4- Compton volume can be
obtained as
\begin{equation}
w_{P}=\sum_{n>n_{0}}w_{n}  \label{wpopov}
\end{equation}%
With the parameter of the theoretical treatments being \ $\gamma =mc\omega /e%
\mathcal{E}=\hbar \omega \mathcal{E}_{c}/mc^{2}\mathcal{E}$, which is the
equivalent of Keldysh parameter, the two important areas $\gamma \ll 1$
(high electric field strength and low frequency, where the adiabatic
non-perturbative tunneling mechanism dominates) and $\gamma \gg 1$(low
electric field strength and high frequency, multiphoton mechanism), lead
respectively to the following number of pairs $N(\tau )$ \cite{Popov1} :

\begin{eqnarray*}
N(\tau ) &=&2^{-3/2}n_{0}^{4}\left( \mathcal{E}/\mathcal{E}_{c}\right) ^{%
\frac{5}{2}}\exp (-\frac{\pi \mathcal{E}_{c}}{\mathcal{E}}(1-\frac{1}{%
2\left( n_{0}\frac{\mathcal{E}}{\mathcal{E}_{c}}\right) ^{2}}))(\omega \tau
/2\pi ),\text{ }\gamma \ll 1 \\
N(\tau ) &\approx &2\pi n_{0}^{3/2}\left( \frac{4\gamma }{e}\right)
^{-2n_{0}}(\omega \tau /2\pi ),\text{\ }\gamma \gg 1
\end{eqnarray*}%
with $\tau $ being the pulse duration and $n_{0}=2mc^{2}/\hbar \omega $. The
first experimental verification of $e^{-}e^{+}$ pair production took place
at SLAC ( E-144 experiment)\cite{Burke}. Highly energetic electrons (maximum
46.6GeV) underwent nonlinear Compton scattering with $n$ laser photons (of $%
\lambda =$527nm, $\omega =2.35eV$ and laser pulses of energy 500mJ ), $%
e+n\omega \rightarrow e^{\prime }+\omega ^{\prime }$ , producing \
backscattered photons $\omega ^{\prime }$ of \ $27-30GeV$\ \ which then
collide with the laser photons to produce $e^{-}$, $e^{+}$ pairs, via the
multiphoton Breit -Wheeler\ mechanism, \ $\omega ^{\prime }+n\omega
\rightarrow e^{-}$ $e^{+}$. The number of positrons measured in 21962 laser
pulses was $175\pm 13$ for a $n=5.1\pm
0.2(statistical)_{-0.8}^{+0.5}(systematic)$ multiphoton order process, a
result that is in very good agreement with the theory. These results have
been shown in \cite{ctm} to be obtained also from Popov's theory.

In a recent paper Avetissian et al \cite{Avetissian} treated $e^{-}e^{+}$
pair production in a standing wave $\mathbf{A}=2\mathbf{A}_{0}\cos \omega t$
formed by oppositely directed laser beams of plane transverse linearly
polarized electromagnetic waves of frequency $\omega ,\;$using a two level
on resonance multiphoton approximation \cite{Avetissian, ctm, tmp}. The main
difference between this approach and the previous one is the condition of
resonance, \ which as shown in the detailed investigation presented in \cite%
{tmp, thesis, ploumis,Tsoha}, can lead to very high production efficiency.
The probability density of pair creation for the case of oscillating
electric field, is given in relation (33) in \cite{Avetissian}$.$This
approximation will be considered in the rest of the paper. Following \cite%
{Avetissian, tmp}, taking the momentum of the created electrons(positrons)
to be $\mathbf{p}=(p_{x}=p\sin \theta ,p_{y}=p\cos \theta ,0)$, where $%
\theta $ is angle between $\mathbf{A}_{0}$ (lying on the Oy axis) and $%
\mathbf{p}$, the probability density $w_{n0}$and the number of pairs created
$N_{0}$ at angle $\theta =0$ (where both maximize) for an $n$th order
process are respectively given by \cite{tmp, ploumis, thesis}

\begin{equation}
w_{n0}=\frac{n\omega }{8\pi ^{2}}f_{n0}^{2}\left( n^{2}\omega
^{2}-4m^{2}\right) ^{\frac{1}{2}}  \label{wNo}
\end{equation}%
\begin{equation}
N_{0}=w_{n0}V\tau =\frac{1}{4\pi ^{2}}\frac{V\tau }{V_{e}}\frac{q\sqrt{%
q^{2}-1}}{m^{2}c^{4}}f_{n0}^{2}  \label{Npo}
\end{equation}

In deriving (\ref{wNo}) and (\ref{Npo}) the resonance condition
\begin{equation}
n=2E/\hbar \omega =2qmc^{2}/\hbar \omega ,\;q\geq 1  \label{res}
\end{equation}

has been used and

\begin{equation*}
f_{n0}=\frac{E}{4p_{y}}(1-\frac{p^{2}}{E^{2}})^{\frac{1}{2}}n\hbar \omega
J_{n}(4\xi \frac{mp}{E\omega }).
\end{equation*}

$\xi \sim \frac{1}{\gamma }$ is the relativistic invariant parameter given by

\begin{equation}
\xi =\frac{mc^{2}\mathcal{E}_{0}}{\hbar \omega \mathcal{E}_{c}},  \label{ksi}
\end{equation}

$\mathcal{E}_{0}\mathcal{\ }$is the amplitude of one incident wave and $\xi
\lesssim 1$ for the approximation to hold. $V_{e}=7.4\times 10^{-59}m^{3}s$
is the 4-Compton volume of the electron, $V\sim \sigma ^{2}l$ is the
interaction volume, $\tau $ is the pulse duration, $\sigma $ is the cross
section radius and $l\ll \lambda $ the electromagnetic wavelength \cite%
{Avetissian, tmp}. Finally since in the Bessel function $J_{n}\left(
x\right) $, $x\simeq n$, the approximation $J_{n}(n\;\sec h\alpha )=(1/\sqrt{%
2\pi n\;\tanh \alpha })$ ($a=sech^{-1}(\frac{2\xi }{q}\left( 1-\frac{1}{q^{2}%
}\right) ^{\frac{1}{2}})$) has to be used giving,
\begin{equation}
f_{n0}=\frac{1}{4}\left( q^{2}-1\right) ^{-\frac{1}{2}}n\hbar \omega \frac{%
\exp \left( n\tanh a-na\right) }{\sqrt{2\pi n\tanh a}}  \label{fNo}
\end{equation}

As shown in \cite{tmp}, relation (\ref{Npo}) leads to very good estimates
for the theory both when higher harmonics are implemented and when the range
of its applicability is examined. For example for a Nd-Yag laser with $%
\omega =1.17eV$, $\lambda =1\mu m$, intensity $1.35\times 10^{22}W/m^{2}$, $%
\tau \sim 10^{-14}s$ , $\sigma \sim 10^{-5}m$, $l=0.1\lambda ,$ the fifth
harmonic can create approximately up to $10^{12}$ pairs for $n\sim 10^{5}$,
provided that the electric field strength does not exceed the order of $%
10^{13}V/m$.

In section two that follows we present the use of a table top laser based
X-ray Free Electron Laser scheme for pair creation and in section three a
scheme analogous to the one used in the E-144 experiment. The aim of this
paper is to emphasize the advantages of using such schemes for the
experimental observation of the on resonance approximation \cite{Avetissian,
tmp}, as they can be realized in well equipped laboratory and can provide
satisfactory results which we will show on our numerical estimations that
will follow. One should mention that the following two setups allow the
increase of electric field strength $\mathcal{E}$ with a simultaneous
increase of $\omega ,$ so that $\xi $ remains $\lesssim 1$, as it is
essential for the validity of the multiphoton on resonance approximation.

\section{\textbf{Pair creation using laser based X -FEL system}}

In this section we investigate the use of what one could call a table top
laser based X-ray Free Electron Laser for electron positron pair creation.
Such system is similar to a standard X -FEL system \cite{xfel} with the main
difference between the two to be the electron beam creation and acceleration
mechanism. The table top XFEL\ utilizes a relativistic electron beam
produced and accelerated by a high intensity ($\sim 10^{20}-10^{21}W/cm^{2}$%
) ultrashort laser beam\cite{Amiranoff, Mangles, Leemans, Malka}. Then the
electron beam propagates through a wiggler system and X- ray beam is
created. Generation of such relativistic electron beam from a laser source
can be achieved by interaction with a solid or gas target and is a wide
field of research. For our purposes we consider a laser system with
intensity of $10^{21}W/cm^{2}$ and pulse duration $20fs$ - $30fs.$ Recent
experiments \cite{Amiranoff, Mangles, Leemans, Malka} have confirmed that
the accelerated electron beam can reach energies $E_{ebeam}$ close to $1GeV$
and charge close to $1nCb$, but is expected that in the near future this
charge will become available. This beam enters an undulator with period $%
\lambda _{u}$ and after the interaction with the magnetic field in it, X-ray
photons of wavelength
\begin{equation}
\lambda =\frac{\lambda _{u}}{2\gamma _{L}^{2}}  \label{lamda}
\end{equation}%
will be produced, where $\gamma _{L}$ is the relativistic factor $(\gamma
_{L}=\frac{E_{beam}}{m_{e}c^{2}})$. The undulator period for our scheme is $%
\lambda _{u}=5mm$ but it can be adjusted depending on the X ray photon
wavelength we wish to produce.

Using (\ref{lamda}) for $E_{ebeam}=1GeV$ \ we obtain X-ray beam wavelength $%
\lambda =0.65nm$,which corresponds to photon energy $\omega =1.909KeV$. Two
opposite directed X-ray beams form a standing wave to a circular spot \ area
of radius $\sigma =100nm$. The interaction 4-volume in which pair creation
takes place is taken as $V=\sigma ^{2}(0.1\lambda )\tau $, where $\tau
=100fs $. We also assume that the conversion efficiency of the electron beam
energy to X-ray photons is about 10\% and given the power of the electron
beam $P_{ebeam}=1\times 10^{13}W$, we obtain the energy of the X-ray beam $%
E_{b}=0.1J$. As in \cite{tmp},with the above parameter values, we firstly
investigate, using (\ref{Npo}) the dependence of pair number $N_{0}$ to $q$
(electron rest mass units). The result is presented in Figure 1, where the
envelope of $N_{o}$($q)$ is plotted, for $\xi =0.9532$. This optimum choice
of $\ \xi $\ (leading to high production efficiency), is dictated by energy
conservation between the X-ray beam energy $E_{b\text{ }}$ and the energy $%
E=2qmc^{2}N_{0}$ of created pairs: one can numerically solve the energy
conservation equation $E_{b}=0.1J=2qmc^{2}N_{0}(\xi ,q)$ for appropriate
range of values of the $q$ parameter (see \cite{tmp} for a detailed analysis
on the choice of $\xi $).

\begin{figure}[tbph]
\includegraphics[width=8.5cm]{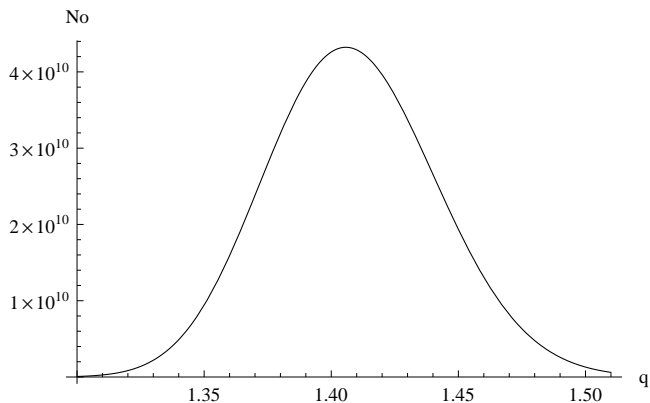}\newline
\caption{Envelope of created e$^{+}$ e$^{-}$ pair number $N_{o}$ vs rest
mass unit $q$ for $\protect\xi =0.9632$ and photon energy $1.909keV.$}
\label{fig1}
\end{figure}
The most probable pair number is the one that corresponds to $q=1.40$ and is
of the order of $10^{10}$ pairs. The multiphoton order of the process as
obtained by (\ref{res}) is $\ n=750$.

Secondly one can also investigate the dependence of $N_{o}$ from $\xi $
(equivalently $\mathcal{E}_{0}$). Using (\ref{Npo}) and the data from Table
1 we present in Figure 2, the dependence of pair number $N_{0}$ from $\xi $
for three $E_{ebeam}$ values of $1GeV$, $400MeV$ and $200MeV$, the last two
being considered in order to estimate and compare pair creation efficiencies
for energies that can be easily created with the recent technology in large
laser facilities. Note that in order to obtain each of the three plots in
Figure 2 the values of ($q$, $n$) are respectively ( 1.41, 750), (1.40,
4742), (1.40,18833).

\begin{figure}[tbph]
\includegraphics[width=7.5cm]{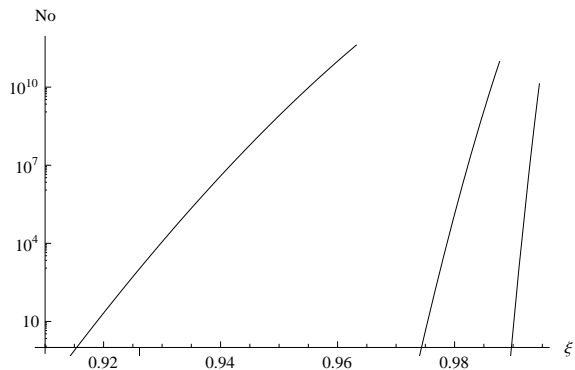}\newline
\caption{Log- plot of the number of created pairs $N_{0}$ vs $\protect\xi $
for electron beam energy $1GeV$ (top curve), $400MeV$ (middle curve) and $%
200MeV$ (bottom)}
\label{fig2}
\end{figure}

Each one of the curves has a different ending point, because of previous
mentioned energy conservation consideration. The choice of plotting vs.
parameter $\xi $ is justifiable, since it depends both from the electric
field strength $\mathcal{E}_{o}$ and photon energy and thus allowing to
investigate the dependence of $\ N_{0}\;$on both of these physical
quantities. By adjusting the photon energy and/or the electric field one can
estimate the number of produced pairs for the corresponding value of $\ \xi $
from figures like Figure 2. Observe the very sensitivity of $N_{0}$ from $%
\xi $ which is a result of resonance approximation(this is due to the
presence of the Bessel function in rel (\ref{Npo})).

\bigskip

\begin{center}
\begin{tabular}{|c|c|c|c|}
\hline
Electron beam & Wavelength of \ X ray & X-ray photon & Maximum Electric field
\\
energy $E_{ebeam}$ & laser beam ($nm$) & energy $(eV)$ & Strength $(V/m)$ \\
\hline
$1GeV$ & 0.65 & 1909 & $3.4\times 10^{15}$ \\ \hline
$400MeV$ & 4.06 & 305 & $5.5\times 10^{14}$ \\ \hline
$200MeV$ & 16.3 & 76 & $1.4\times 10^{14}$ \\ \hline
\end{tabular}

\bigskip Table 1. Typical parameters for obtaining Figure 2

The electron beam charges considered for these energies case are
respectively 1nCb for 1GeV , 0.4 nCb for 400MeV and 0.2 nCb for 200MeV. The
estimates of Figure 2 show that even if we choose to lower the electron beam
energy, we can achieve efficient electron positron pair production ( of the
order of 10$^{11}$for 1GeV , 10$^{10}\;$for 400 and 200 MeV) and strongly
suggest the possible use of this scheme for experimental observation of
resonance approximation.
\end{center}

\section{\textbf{Pair creation using a configuration analogous to the E144
experiment.}}

The second scheme proposed is based on the E144 experiment at SLAC \cite%
{Burke}.It consists of two steps. On the first step a relativistic electron
beam is produced and accelerated by an ultrashort laser beam of photon
energy $1eV$. On the second step the relativistic electron beam interacts
with the main high intensity (10$^{20}\;W/cm^{2}$) highly focused laser
beam. On the electron's frame the electric field intensity of the laser beam
is thus increased due to the relativistic factor while the focusing
guarantees that the condition $\mathcal{E}^{2}-H^{2}>0,$ for pair creation
is satisfied. On the electrons reference frame the electric field $\mathcal{E%
}^{\ast }$ and the photon energy $E_{photons}^{\ast }\;$are given by

\begin{eqnarray}
\mathcal{E}^{\ast } &=&\gamma _{L}\;\mathcal{E}_{laserlab}  \label{trelf} \\
E_{photons}^{\ast } &=&\gamma _{L}\;E_{photons}  \label{trphot}
\end{eqnarray}

where $\gamma _{_{L}}=\frac{E_{ebeam}}{m_{e}c^{2}}$, $E_{ebeam}$ being the
electron beam energy.

For $E_{ebeam}=1GeV,\;$we obtain from (\ref{trphot}) $E_{photons}^{\ast
}=1.956\;keV$ , which corresponds to $\lambda =$ $0.1nm$ and $\mathcal{E}%
^{\ast }\sim 10^{15}V/m$. We will apply the same analysis as in the section
two , to estimate the efficiency of resonance approximation for this scheme.
Using (\ref{Npo}) where the interaction 4-Volume $V=\sigma ^{2}(0.1\lambda
)\tau $, $\lambda =0.1nm,\;\sigma =50\mu m\;$and $\tau =10^{-14}s$ ,we
present in Figure 3 the envelope of the created pair number $N_{0}$ as a
function of $q$ for $\xi =0.9632$ ($\xi $ is obtained with same
considerations as before). $\;$

\begin{figure}[tbph]
\includegraphics[width=8.5cm]{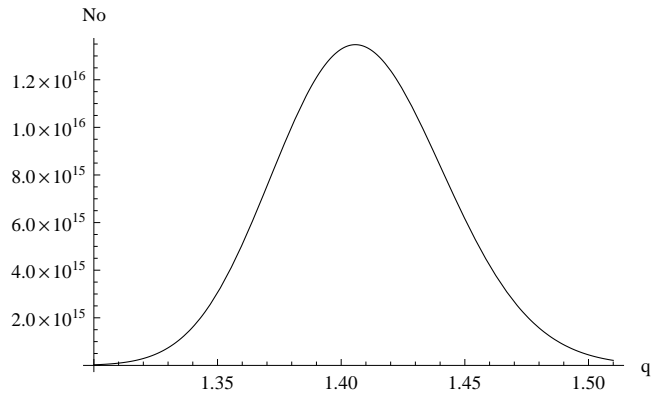}\newline
\caption{Envelope of created e$^{+}$ e$^{-}$ pair number $N_{o}$ vs rest
mass unit $q$ for $\protect\xi =0.9632$ and photon energy $1.956keV.$}
\label{fig3}
\end{figure}

The most probable process corresponds to $q=1.41$ and can produce up to the
order of $10^{16}$ pairs. The multiphoton order of this process is $n=735$.
We also examine the efficiency of pair creation for two more energies of the
electron beam, $400MeV$ and $200MeV.$ As in section two, using formula (\ref%
{Npo}) and the data from Table 2 we present in Figure 4, the dependence of
pair number $N_{0}$ from $\xi $ for three $E_{ebeam}$ values of $1GeV,$ $%
400MeV$ and $200MeV$ with corresponding values ($q$, $n$) for the later two
energies as (1.41, 1844) and (1.40, 3686)

\begin{figure}[tbph]
\includegraphics[width=8.5cm]{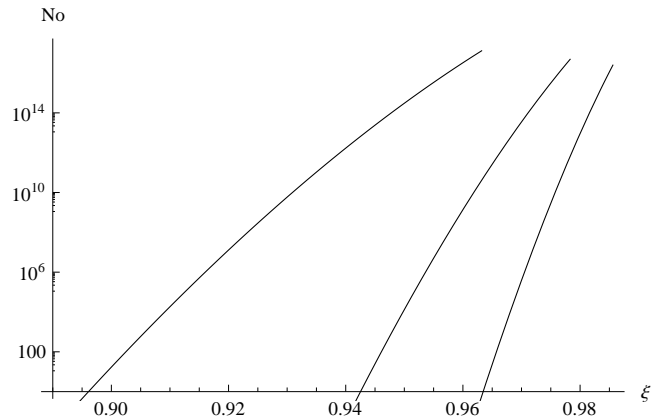}\newline
\caption{Log- plot of the number of created pairs $N_{0}$ vs $\protect\xi $
for electron beam energy $1GeV$ (top curve), $400MeV$ (middle curve) and $%
200MeV$ (bottom)}
\label{fig4}
\end{figure}

\begin{center}
\begin{tabular}{|c|c|c|}
\hline
Electron beam & X-ray photon & Electric field \\
energy $E_{e-beam}$ & energy $(eV)$ & Strength $(V/m)$ \\ \hline
$1GeV$ & 1909 & $4.7\times 10^{15}$ \\ \hline
$400MeV$ & 782 & $7.8\times 10^{14}$ \\ \hline
$200MeV$ & 391 & $1.9\times 10^{14}$ \\ \hline
\end{tabular}

\bigskip Table 2. Typical parameters for obtaining Figure 4

The estimates of Figure 4 of this scheme strongly suggest the possible use
of this scheme for experimental observation of resonance approximation ( of
the order of 10$^{16}$for 1GeV , 10$^{15}\;$for 400 and 200 MeV).
\end{center}

\bigskip

\section{Conclusion}

The above analysis aims to support the experimental verification of e$%
^{+}e^{-}$ pair creation using the on resonance multiphoton approximation by
presenting two possible schemes. One is based on using a table-top X-FEL
laser system and the second is analogous to the E144 experiment. The
numerical estimates presented that a sufficient number of pairs can be
produced by both schemes that can reach up to the order of \ 10$^{11}$ pairs
for the table top X-FEL and 10$^{15}$ for the E144 like scheme. An important
fact in both proposed schemes is that they use laser based modern technology
for production and acceleration of electron beam and consequently they do
not need any large acceleration facilities. Moreover, since the existing
laser laboratories have this kind of technology, the required experimental
values of the parameters needed for pair production can be achieved. Our
choice to carry out our numerical estimates with $E_{ebeam}\ $ no greater
than 1GeV is thus justified, as electron beams with energies like 400 MeV or
200\ MeV can be created quite easily. Besides the above advantages of the
two schemes, the laser beam parameters can be easily adjusted to change the
electron beam energy, in order to operate on resonance. Moreover the
criterion for the choices of the values of the parameters used, is that they
should be close to typical values rather than extreme ones, in order to
maintain our estimates as realistic as possible. Consequently our analysis,
suggests that both of the schemes presented are promising for a future
experimental verification of the on resonance pair production, especially
when new powerful laser systems like X-FEL\cite{xfel}, HIPER\cite{Hiper} and
ELI\cite{ELI} are under development.


\begin{thebibliography}{99}
\bibitem{Klein} O. Klein, Z. Phys., \textbf{53}, 157 (1929); F. Sauter, Z.
Phys. \textbf{69}, 742 (1931); W. Heisenberg, H. Euler, Z. Phys. \textbf{98}%
, 718 (1936).

\bibitem{Schwinger} J. W. Schwinger, Phys. Rev., \textbf{82}, 664 (1951).

\bibitem{Brezin} E. Brezin and C. Itzykson, Phys. Rev. D \textbf{2}, 1191
(1970).

\bibitem{Troup} G.J. Troup and H.S. Perlman, Phys. Rev. D \textbf{6}, 2299
(1972).

\bibitem{greiner} W. Greiner, B. Muller, J. Rafelski, `Quantum
Electrodynamics of Strong Fields', Springer -- Verlag, Berlin, 1985.

\bibitem{Popov1} V.S. Popov, JETP Lett. \textbf{13}, 185 (1971); Sov. Phys.
JETP \textbf{34}, 709 (1972); Sov. Phys. JETP \textbf{35}, 659 (1972); V.S.
Popov and M. S. Marinov, Sov. J. Nucl. Phys.\textbf{16}, 449 (1973) ; JETP
Lett. \textbf{18}, 255 (1974); Sov. J. Nucl. Phys., \textbf{19}, 584 (1974).

\bibitem{Nikishov2} A. I. Nikishov, Nucl. Phys. B\textbf{21}, 346 (1970).

\bibitem{niki} N.B. Narozhnyi and A. I. Nikishov, Sov. J. Nucl. Phys.\textbf{%
11}, 596 (1970); \ Sov. Phys. JETP, \textbf{38}, 427 (1974).

\bibitem{Ringwald} A. Ringwald, Phys. Let. B \textbf{510}, 107 (2001).

\bibitem{popov2} V.S. Popov, Phys. Let. A\textbf{298}, 83 (2002).

\bibitem{mourou} M. Perry and G. Mourou, Science \textbf{264}, 917 (1994)

\bibitem{chen} P. Chen and C. Pellegrini, in Quantum Aspects of Beam
Physics, Proc.15th Advanced ICFA Beam Dynamics Workshop, Monterey, Cal., 4-9
Jan 1998 (World Scientific, Singapore, 1998) p. 571.

\bibitem{tajima} P. Chen and T. Tajima, Phys. Rev. Lett. \textbf{83}, 256
(1999).

\bibitem{mourou2} T. Tajima and G. Mourou, Phys. Rev. STAccel.Beams 5,
031301 (2002)

\bibitem{bulanov} S. S. Bulanov,N.B. Narozhny,V.D. Mur, and V.S. Popov,
Phys. Lett. A 330, 1 (2004)

\bibitem{blaschke} D. B. Blaschke, A.V. Prozorkevich, C. D. Roberts, S. M.
Schmidt and S. A. Smolyansky, Phys. Rev. Lett, 96, 140402 (2006).

\bibitem{xfel} XFEL-Technical Disign Report (2006) Publisher: DESY XFEL
Project Group, European XFEL Project Team, Deutsches Elektronen-Synchrotron,
Member of the Helmholtz Association, Notkestrasse 85, 22607 Hamburg, Germany
ISBN 3-935702-17-5

\bibitem{Burke} D.L. Burke et. al., Phys. Rev. Let., \textbf{79}, 1626
(1997).

\bibitem{Avetissian} H. K. Avetissian, A. K. Avetissian, G. F. Mkrtchian and
Kh. V. Sedrakian, Phys. Rev. E \textbf{66}, 016502 (2002).

\bibitem{ctm} C. Kaberidis, I. Tsohantjis and S. Moustaizis in `Frontiers of
Foundamental and Computational Physics' Udine, Italy, 26-29 September 2004,
Sidharth B.G, Honsell F., de Angelis A. (Eds.) 2005 pp. 279-283

\bibitem{tmp} I. Tsohantjis, S. Moustaizis and I. Ploumistakis, Physics
Letters B 650 249(2007)

\bibitem{thesis} I. Ploumistakis, Master Thesis Technical University of
Crete, (2007)

\bibitem{ploumis} I. Ploumistakis, I. Tsohantjis and S. Moustaizis, "New
approaches on Laser Vacuum Breakdown for Pair Creation", 35th EPS Conference
on Plasma Phys. Hersonissos, 9 - 13 June 2008 ECA Vol.32D, P-1.123 (2008)

\bibitem{Tsoha} I.Tsohantjis, S. D. Moustaizis, I. Ploumistakis, "Pair
creation from vacuum in the presence of ultra-intense laser beams", 35th EPS
Conference on Plasma Phys. Hersonissos, 9 - 13 June 2008 ECA Vol.32D,
O-4.041 (2008)

\bibitem{Malka} V. Malka, J. Faure, Y.Glinec, A. Pukhov, J.Rousseau Phys. of
Plasmas 12, 056702 (2005)

\bibitem{Amiranoff} F. Amiranoff et.al Phys. Rev. Lett. 81, 995 (1998)

\bibitem{Leemans} W.P Leemans, et.al Nature Physics 2, (2006)

\bibitem{Mangles} S.P.D. Mangles, et.al Nature 431, 535 (2004)

\bibitem{Hiper} http://www.hiper-laser.org

\bibitem{ELI} http://www.extreme-light-infrastructure.eu
\end{thebibliography}
\end{document}